\documentstyle[12pt,a4]{article}

\newcommand{\hl}{\hline}
\newcommand{\ol}{\overline}
\newcommand{\VEV}[1]{\left\langle #1 \right\rangle}

\newcommand{\mybar}[1]%
   {\kern 0.8pt\overline{\kern -0.8pt#1\kern -0.8pt}\kern 0.8pt}
\newcommand{\drawsquare}[2]{\hbox{%
\rule{#2pt}{#1pt}\hskip-#2pt%
\rule{#1pt}{#2pt}\hskip-#1pt%
\rule[#1pt]{#1pt}{#2pt}}\rule[#1pt]{#2pt}{#2pt}\hskip-#2pt%
\rule{#2pt}{#1pt}}%
\newcommand{\f}{\raisebox{-.5pt}{\drawsquare{6.5}{0.4}}}
\newcommand{\fb}{\overline{\f}}
\newcommand{\asym}{\raisebox{-3.5pt}{\drawsquare{6.5}{0.4}}\hskip-6.9pt
  \raisebox{3pt}{\drawsquare{6.5}{0.4}}}
\newcommand{\asymb}{\overline{\asym}}

\begin{document}
\begin{titlepage}
 \begin{flushright}
    KUNS-1544\\
    HE(TH) 98/18\\
    hep-th/9811119
  \end{flushright}
  
  \vspace*{4ex}
  \begin{center}
    {\Large\bf
      Duality between simple-group gauge theories\\[2mm]
      and some applications}
    \vspace{7ex}

    Takayuki Hirayama\footnote{e-mail:
      hirayama@gauge.scphys.kyoto-u.ac.jp}
    and
    Koichi Yoshioka\footnote{e-mail:
      yoshioka@gauge.scphys.kyoto-u.ac.jp}
    \vspace{2ex}

    {\it Department of Physics, Kyoto University, Kyoto 606-8502,
      Japan}
  \end{center}
  \vspace*{9ex}

%%%%%%%%%%%%%%%%%%%%%%%%%%%%%%%%%%%%%%%%%%%%%%%%%%%%%%%
%%%%%%%%%%%%%%%%%%%%%%%%%%%%%%%%%%%%%%%%%%%%%%%%%%%%%%%
%%%%%--------------y   abstract   z-------------%%%%%
  \begin{abstract}
    In this paper we investigate N=1 supersymmetric gauge theories
    with a product gauge group. By using smoothly confining dynamics, we
    can find new dualities which include higher-rank tensor fields,
    and in which the dual gauge group is simple, not a product. Some
    of them are dualities between chiral and non-chiral gauge
    theories. We also discuss some applications to dynamical
    supersymmetry breaking phenomena and new confining theories with
    a tree-level superpotential. \\[4mm]
  \end{abstract}
\end{titlepage}

\setcounter{footnote}{0}

%%%%%%%%%%%%%%%%%%%%%%%%%%%%%%%%%%%%%%%%%%%%%%%%%%%%%%%
%%%%%%%%%%%%%%%%%%%%%%%%%%%%%%%%%%%%%%%%%%%%%%%%%%%%%%%
%%%%%---------------y  section  z---------------%%%%%
\section{Introduction}

${\cal N}=1$ supersymmetric gauge theories in four dimensions are very
attractive and have been intensively investigated. Recent developments
in understanding non-perturbative effects reveal the strong dynamics of
supersymmetric gauge theories. It has been applied to
constructing various models beyond the standard model, dynamical
supersymmetry breaking models \cite{DSB}, composite models of quark
and lepton \cite{composite}, and so on. Especially, phenomena of
duality in supersymmetric gauge theories appear in many different
gauge and superstring theories. So, the study of duality itself may be
important.

We know that ${\cal N}=1$ supersymmetric gauge theories have
various phases of dynamics in the infrared region. For example,
consider the $SU(N)$ gauge theory with $N_F$ flavors of fundamental
matter fields \cite{seiberg}. For $N_F<N$ the theory has no stable
supersymmetric vacuum (runaway behavior) because the
Affleck-Dine-Seiberg superpotential is dynamically generated. When
$N_F=N$ the theory confines with chiral symmetry breaking
(i-confinement), and when $N_F=N+1$ the theory confines without chiral
symmetry breaking (s-confinement). In case $N_F>N+1$ the theory has
a dual description which has the same infrared behavior as the
original theory.

So far, various models have been known to have dual
descriptions. Among these we know several examples of duality in
theories with matters in the higher-rank tensor representation. When
using the deconfinement method \cite{berkooz} to expand these
higher-rank tensor fields, in general, the gauge group of the dual
theory is the product group \cite{berkooz}--\cite{terning}. On the other
hand, the Kutasov-Schwimmer-type dualities \cite{kutasov,kutasov2} and
the dualities in spin gauge theories with matters in spinorial
representation \cite{pouliot2}--\cite{cho} have simple dual gauge
groups. It is, however, difficult to find these types of dualities
because no systematic method has been discovered as of yet. In this
paper, by applying the deconfinement method, we propose a systematic
way to find examples of such duality. With this method, we can
construct many examples of duality including chiral to non-chiral
dualities and a duality for the exceptional $G_2$ gauge theory.

This paper is organized as follows. In Section 2 we shortly review the
deconfinement method and discuss our idea for finding duality. In
Section 3 we show several examples and check the consistency in an
example, and we show some more examples and applications to dynamical
supersymmetry breaking and confining theories in Section 4. Section 5
is devoted to summary and conclusion.

%%%%%%%%%%%%%%%%%%%%%%%%%%%%%%%%%%%%%%%%%%%%%%%%%%%%%%%
%%%%%%%%%%%%%%%%%%%%%%%%%%%%%%%%%%%%%%%%%%%%%%%%%%%%%%%
%%%%%---------------y  section  z---------------%%%%%
\section{The idea}
\setcounter{equation}{0}

In this section we describe our method for finding duality by using a
variant of the deconfinement method. In the deconfinement method, we
apply the confining dynamics to expanding higher-rank tensor fields in
order to obtain the deconfined theory with only fundamental
representation matters. We first summarize the sorts of confining
theories.

In the absence of tree-level superpotential, there are four types of
confining theory which are formally called s-confinement \cite{csaki},
i-confinement \cite{grinstein}, c-confinement \cite{grinstein}
and affine-confinement \cite{dotti1,dotti2}, respectively. The
s-confinement is defined as a theory for which the low-energy
effective theory is described by gauge invariant operators everywhere
in the moduli space. In addition, the structure of quantum moduli
space is the same as that of the classical theory. Therefore the
effective superpotential
is certainly generated to reproduce the classical constraints as
equations of motion. In i- (c-) confining theories the classical
moduli is deformed by quantum mechanics. The global symmetry is broken
in i-confining theories. In c-confining theories the origin of moduli
space is allowed and global symmetries are not broken at that
point. The affine-confinement is the confining theory without any
constraints between the composite states which describe the low-energy
dynamics. We summarize the types of confining theories in Table
\ref{tab:conf}.
\begin{table}[htbp]
  \def\arraystretch{1.1}
  \begin{center}
    \leavevmode
    \begin{tabular}{|c|l|} \hline
      type & \multicolumn{1}{c|}{example} \\ \hline\hline
      s-confinement & $SU(N)$ with $(N+1)\, \f$ \\ \hline
      i-confinement & $SU(N)$ with $N\, \f$ \\ \hline
      c-confinement & $SU(4)$ with $3\, \asym + (\f+\fb)$ \\ \hline
      affine-confinement & $SO(N)$ with $(N-4)\, \f$ \\ \hline
    \end{tabular}
    \caption{The classification of the confining theories.}
    \label{tab:conf}
  \end{center}
\end{table}

%%%%%%%%%%%%%%%%%%%%%%%%%%%%%%%%%%%%%%%%%%%%%%%%%%%%%%%
\subsection{The deconfinement method}

Before we discuss our idea, we shortly review the deconfinement
method proposed by Berkooz \cite{berkooz}. The deconfinement method is
used to find the dual theories (the magnetic theories) of the theories
(the electric theories) containing the two index tensor representation
(antisymmetric, symmetric or adjoint representation) matters.

The electric theory is expanded into the theory (the deconfined theory)
as the two index tensor fields replaced by a composite operator. For
example the $SU(N)$ symmetric tensor $S^{ab}$ can be replaced by
\begin{eqnarray}
  S^{ab} &\rightarrow& [x^ax^b]
\end{eqnarray}
where the field $x^a$ belongs to the vector representation of
$SO(N+5)$ gauge symmetry which now confines without chiral symmetry
breaking. Here and throughout this paper, we write composite states as
[$\cdots$]. The matter content and superpotential are shown in Table
\ref{tab:deconf}. In this table, we omit the $SU(N)$ gauge anomaly
which is usually compensated by more antifundamental fields.
\begin{table}[htbp]
  \begin{center}
    \leavevmode
    $\begin{array}{|c|cc|} \hline
      & SU(N) & SO(N+5) \\ \hline
      x^a_i & \f  & \f \\
      y_i   &  1  & \f \\
      z_a   & \fb &  1 \\
      u     &  1  &  1 \\ \hline
    \end{array}$ \\ \vspace*{2ex}
    $W= xyz + y^2u$
  \end{center}
  \caption{The deconfined theory for $SU(N)$ theory with a symmetric
    tensor field.}
  \label{tab:deconf}
\end{table}
When this $SO(N+5)$ gauge group confines before the $SU(N)$ gauge
coupling becomes strong, i.e.,\ $\Lambda_{SU} \ll \Lambda_{SO}$
(the dynamical scale $\Lambda_{SU}$ ($\Lambda_{SO}$) means that it is
the scale when $SO$ ($SU$) gauge coupling turns off), no dynamical
superpotential is generated. The resultant $SU(N)$ theory is described
by the $SO(N+5)$ gauge singlet composite states,
\begin{eqnarray}
  [x^ax^b],\quad [x^ay],\quad [yy]\,.
\end{eqnarray}
The superpotential makes the fields $[x^ay]$ and $[yy]$ massive. We
integrate out these massive modes by using their equations of motion
and then we obtain the $SU(N)$ gauge theory with a symmetric tensor
field $S^{ab}$ (the electric theory).

A way to find the magnetic description of this electric theory is
apparent. The holomorphy of moduli space tells that the
equivalence of the infrared limit is valid for all value of
$\Lambda_{SU} /\Lambda_{SO}$ \cite{product}. So, when taking the
dual of the $SU(N)$ gauge symmetry in the above deconfined theory at
first, i.e.,\ in the case of $\Lambda_{SU} \gg \Lambda_{SO}$, we
can obtain a magnetic description of the electric theory. Moreover,
if we dualize the $SO(N+5)$ gauge symmetry in this magnetic theory, we
obtain the second magnetic theory. In this way we can find the
sequence of the dual magnetic theories.
We can easily check the 't Hooft anomaly matching conditions and the
correspondence of the gauge invariant operators between the electric
and magnetic theories because they have been confirmed between the
deconfined theory and each theory.

%%%%%%%%%%%%%%%%%%%%%%%%%%%%%%%%%%%%%%%%%%%%%%%%%%%%%%%
\subsection{The idea}

With the deconfinement method described in the last subsection, one
can find duality between the electric theory including the two index
tensor representation matters and the magnetic theory whose gauge
group is necessarily a product. We display this procedure in Table
\ref{tab:chart-deconf}.
\begin{table}[htbp]
  \begin{center}
    \leavevmode
    \fbox{$G_1 \times G_2$ (duality $\times$ confinement)} \\[1.5mm]
    the deconfined theory\\
    $\downarrow\hspace*{24ex}\downarrow$\\[2mm]
    \begin{tabular}{c}
      \fbox{$G_1$ with tensors}\\[2mm]
      the electric theory
    \end{tabular}\hspace*{6ex}
    \begin{tabular}{c}
      \fbox{$\widetilde G_1 \times G_2$}\\[2mm]
      the magnetic theory
    \end{tabular}
    \caption{The deconfinement method.}
    \label{tab:chart-deconf}
  \end{center}
\end{table}
The label ``duality $\times$ confinement'' means that when the $G_2$
gauge coupling turns off the theory has other descriptions which
describe the same infrared physics, and when $G_1$ gauge coupling
turns off the theory confines.
Therefore if we want to obtain dualities in which both gauge groups
are simple, we take the deconfined theory with the label ``confinement
$\times$ confinement'' (see Table \ref{tab:ours}).
\begin{table}[htbp]
  \begin{center}
    \leavevmode
    \fbox{$G_1 \times G_2$ (confinement $\times$ confinement)}\\[1.5mm]
    the deconfined theory\\
    $\downarrow\hspace*{24ex}\downarrow$\\[2mm]
    \begin{tabular}{c}
      \fbox{$G_1$ with tensors}\\[2mm]
      the electric theory
    \end{tabular}\hspace*{6ex}
    \begin{tabular}{c}
      \fbox{$G_2$ with tensors}\\[2mm]
      the magnetic theory
    \end{tabular}
    \caption{Our method.}
    \label{tab:ours}
  \end{center}
\end{table}

Let us discuss whether the dynamically generated superpotential
changes when global symmetries are gauged and whether we can
expect that there is a dual description for the electric (magnetic)
theory. We comment on these two points below and will check more
non-trivially for some examples in detail in the next section.

The matter content of the deconfined theory can be generally written
as in Table \ref{tab:general}. Here we omit the non-abelian global
flavor symmetries. $R_i$ and $r_i$ ($i=1,\cdots , l$) denote the
representations of the field $\Phi_i$ under the gauge group $G_1$ and
$G_2$\@. Each $U(1)$ flavor symmetry has gauge anomaly, so both the
dynamical scales ($\Lambda_1^{b_1}, \Lambda_2^{b_2}$) have nonzero
charge under these symmetries. The exponents $b_1$ and $b_2$ are the
coefficients of the one-loop beta function of $G_1$ and $G_2$ gauge
couplings. The $U(1)_R$ is the R-symmetry in ${\cal N}=1$
supersymmetric theories. We now take the deconfined theory with the
label "s-confinement $\times$ s-confinement", so the $U(1)_R$ charges
of these dynamical scales are $-2$.
\begin{table}[htbp]
  \begin{center}
    \leavevmode $
    \begin{array}{|c|cc|cccc|c|} \hline
      & G_1 & G_2 & U(1)_1 & U(1)_2 & \cdots & U(1)_l & U(1)_R \\ \hline
      \Phi_1 & R_1 & r_1 & 1 & 0 & \cdots & 0 & 0 \\
      \vdots&\vdots&\vdots&\vdots&\vdots&\ddots&\vdots&\vdots\\
      \Phi_l & R_l & r_l & 0 & 0 & \cdots & 1 & 0 \\ \hline
      \Lambda^{b_1}_1 &   &   & \mu(R_1) & \mu(R_2) & \cdots &
      \mu(R_l) & -2 \\
      \Lambda^{b_2}_2 &   &   & \mu(r_1) & \mu(r_2) & \cdots &
      \mu(r_l) & -2 \\ \hline
    \end{array} $
    \caption{The matter content and the charge assignment of the
      deconfined theory. ($\mu(R)$ is the Dynkin index of
      representation $R$.)}
    \label{tab:general}
  \end{center}
\end{table}
We take the tree-level superpotential
\begin{eqnarray}
  W &=& 0.
\end{eqnarray}

We now consider the case for $\Lambda_1 \ll\Lambda_2$ and examine the
effective theory near below the scale $\Lambda_2$, i.e.,\ we now
take into account the effect of the $G_2$ gauge dynamics. Using the
symmetries and considering the classical limit, the superpotential of
the electric $G_1$ theory takes the same form as that when the $G_1$
gauge coupling turns off;
\begin{eqnarray}
  W &=& \frac{1}{\Lambda_2^{b_2}}\left(
    \Phi_1^{\mu(r_1)} \cdots \Phi_l^{\mu(r_l)}\right).
\end{eqnarray}
Therefore it is found that the superpotential does not change even
when flavor symmetries are gauged. It should be noted that from the
argument of symmetries and the classical limit, the terms which
include the scale $\Lambda_1$ are also allowed. However, these terms
are important only when we consider the low-energy behavior of the
electric $G_1$ theory. A similar analysis can be made in the magnetic
$G_2$ theory in the case of $\Lambda_1\gg\Lambda_2$\@.

The ${\cal N}=1$ supersymmetric asymptotically free gauge theories
without classical superpotential are roughly classified into three
types of theories: runaway, confinement, and duality. Using symmetries
and taking the classical limit, it is found that the relationship
between these types and the Dynkin index constraints arises. The
relation is represented in terms of the sum of indices of matter
fields $\sum \mu_i$ and the index of the adjoint representation
$\mu_{G}$ (see Table \ref{dynamics}).
\begin{table}[htbp]
  \begin{center}
    \leavevmode
    \begin{tabular}{|l|l|} \hline
      \multicolumn{1}{|c|}{type} & \multicolumn{1}{c|}{index
        constraint} \\ \hline\hline
      runaway & $\sum \mu_i<\mu_{G}$ \\ \hline
      confinement & $\sum \mu_i=\mu_{G},\, \mu_{G}+2$ \\ \hline
      duality & $\sum \mu_i \geq \mu_{G}+4$  \\ \hline
    \end{tabular}
    \caption{The rough classification of ${\cal N}=1$ gauge dynamics
        in the absence of the tree-level superpotential.}
      \label{dynamics}
  \end{center}
\end{table}

Since we take a deconfined theory with the label ``s-confinement
$\times$ s-confinement,'' the sum of indices of each theory is equal
to $\mu_{G}+2$\@. We now want to study how the index of the $G_1$
theory changes as a consequence of the $G_2$ confining
dynamics. Applying the results in Ref. \cite{manohar}, we obtain
the following expression for the $G_1$ index after the confinement,
\begin{eqnarray}
  \sum_{i'\in G_1}\mu_{i'} &\geq& \sum_{i\in G_1}\mu_i
  +\sum_{i\in G_2}\mu_i -\mu_{G_2} \label{index}\\
  &=& (\mu_{G_1}+2) +2\,,
\end{eqnarray}
where $i'$ on the left-hand side denotes the kind of fields in the
low-energy
$G_1$ theory after $G_2$ confinement. Note that the inequality
(\ref{index}) is valid only in confining theories except for
affine-confining theories. From this we expect that the electric
theory has a dual which describes the same infrared behavior. We can
obtain the magnetic theory in another limit $\Lambda_1 \gg
\Lambda_2$. In this way we obtain dualities between the electric and
magnetic theories.

{}From the above expression for the Dynkin index, it can be easily seen
that we can construct other types of low-energy $G_1$ gauge theories by
taking $G_2$ gauge theories with other indices. We will show such
examples in subsection 4.3.

%%%%%%%%%%%%%%%%%%%%%%%%%%%%%%%%%%%%%%%%%%%%%%%%%%%%%%%
%%%%%%%%%%%%%%%%%%%%%%%%%%%%%%%%%%%%%%%%%%%%%%%%%%%%%%%
%%%%%---------------y  section  z---------------%%%%%
\section{Example and consistency check}
\setcounter{equation}{0}
\setcounter{table}{0}

In this section, we present several examples of the new dual pair
which
are obtained by the method explained in the previous section. We can
obtain these examples easily since the s-confining theories without a
tree-level superpotential have been completely studied
\cite{csaki}. The massless composite fields are often in higher-rank
tensor representation under the flavor symmetries. Then we have
dualities with fields in higher-rank tensor representation. In
addition, it sometimes occurs that both electric and magnetic theories
have already been known to have other dual descriptions in the absence
of the superpotential. In that case, we can check that these pairs of
theories are indeed connected by duality transformations and describe
the same low-energy dynamics.

The first example is a theory based on the gauge group of $SU(2N)$
and $Sp(2N)$\@. We take the following theory
\begin{table}[htbp]
  \begin{center}
    \leavevmode $
\begin{array}{|c|cc|ccccc|} \hline
  & SU(2N) & Sp(2N) & SU(4) & SU(2N+1) & U(1)_1 & U(1)_2 & U(1)_R \\
  \hline
  Q  &  \f  &  \f  &  1  &  1  &  1  &  0  & \frac{1}{N} \\
  Q' &  1  &  \f  &  \f  &  1  & -\frac{N}{2} &  0  &  0  \\
  Q'' &  \f  &  1  &  1  &  1  &  1  &  2N+1  &  0  \\
  \overline{Q} &  \fb  &  1  &  1  &  \f  &  -1  &  -1  &  0 \\ \hline
\end{array} $
    \caption{The deconfined theory.}
\label{tab:exa-deconf}
  \end{center}
\end{table}
with the tree-level superpotential $W=0$ as the deconfined theory
(Table \ref{tab:exa-deconf}). Both the $SU(2N)$ and $Sp(2N)$ theories
are supposed to confine \cite{s-conf} at the scales of $\Lambda_1$ and
$\Lambda_2$, respectively. First, let us consider the case for
$\Lambda_1 \ll \Lambda_2$, in which case the strong coupling dynamics
of the $Sp(2N)$ theory is important below the scale of
$\Lambda_2$. The resultant confined theory, which we call the electric
theory, is the $SU(2N)$ gauge theory, and has the following matter
content and the dynamically generated superpotential (Table
\ref{tab:exa-ele});
\begin{table}[htbp]
  \def\arraystretch{1.1}
  \begin{center}
    \leavevmode $
\begin{array}{|c|c|ccccc|} \hline
  & SU(2N) & SU(4) & SU(2N+1) & U(1)_1 & U(1)_2 & U(1)_R \\ \hline
  V=[QQ] &  \asym  &  1  &  1  &  2  &  0  & \frac{2}{N} \\
  V'=[QQ'] &  \f  &  \f  &  1  & 1-\frac{N}{2} &  0  & \frac{1}{N} \\
  V''=[Q'Q'] &  1  &  \asym  &  1  &  -N  &  0  &  0  \\
  Q'' &  \f  &  1  &  1  &  1  &  2N+1  &  0 \\
  \overline{Q} &  \fb  &  1  &  \f  &  -1  &  -1  &  0  \\ \hline
\end{array} $
    \caption{The electric theory.}
    \label{tab:exa-ele}
  \end{center}
\end{table}
\begin{eqnarray}
  W = \frac{1}{\Lambda_2^{2N+1}} \left( V^NV''^2 +V^{N-1}V'^2V''
    +V^{N-2}V'^4 \right).
  \label{exa-ele-W}
\end{eqnarray}
This electric theory surely has the index which satisfies $\sum \mu_i
-\mu_G > 2$. Therefore we expect there to be a dual description. As
mentioned earlier, a dual description of this electric theory can be
obtained when one considers another limit of $\Lambda_1 \gg
\Lambda_2$\@. In this limit, the $SU(2N)$ gauge group in the
deconfined theory smoothly confines and the field content of the
resultant magnetic theory is given in Table \ref{tab:exa-mag}.
\begin{table}[htbp]
  \begin{center}
    \leavevmode $
\begin{array}{|c|c|ccccc|} \hline
  & Sp(2N) & SU(4) & SU(2N+1) & U(1)_1 & U(1)_2 & U(1)_R \\ \hline
  M=[Q\overline{Q}] &  \f  &  1  &  \f  &  0  &  -1  & \frac{1}{N} \\
  M'=[Q''\overline{Q}] &  1  &  1  &  \f  &  0  &  2N  &  0  \\
  B=[Q^{2N}] &  1  &  1  &  1  &  2N  &  0  &  2  \\
  B'=[Q^{2N-1}Q''] &  \f  &  1  &  1  &  2N  & 2N+1 & 2-\frac{1}{N} \\
  \overline{B}=[\overline{Q}^{2N}] &  1  &  1  &  \fb  &  -2N  &  -2N
  &  0 \\
  Q' &  \f  &  \f  &  1  & -\frac{N}{2} &  0  &  0  \\ \hline
\end{array} $
    \caption{The magnetic theory.}
    \label{tab:exa-mag}
  \end{center}
\end{table}
This theory has a dynamically generated superpotential
\begin{eqnarray}
  W = \frac{1}{\Lambda_1^{4N-1}} \left( M^{2N}M' +BM'\overline{B}
    +B'M\overline{B} \right).
  \label{exa-mag-W}
\end{eqnarray}
These two theories based on simple gauge groups are certainly dual
in the senses that the moduli space of vacua is identical in a
consequence of the holomorphy of couplings in the moduli space, and
the 't Hooft anomaly matching conditions are trivially satisfied
between the electric and magnetic theories because this is guaranteed
to work originally between the deconfined theory and each theory.

One more support of duality is the consistency against
deformations added in both theories. Especially, in the above case,
two theories have other dual descriptions obtained from the known
dualities, in which the consistency for deformations has already been
shown. Therefore, we can corroborate the above obtained duality by
applying the known dualities to each theory.

First, we take a dual of the electric theory. Though the dual
description of $SU(2N)$ theory with an antisymmetric tensor is known
only in the case of the critical number of flavors by using the
deconfinement method, a dual description of the case of five flavors
has
been discussed by Terning \cite{terning} in terms of a theory with
gauge group $SU(2)\times SU(2)$\@ (see Table \ref{tab:dualofle}). When
applying this dual
transformation, some terms in the superpotential (\ref{exa-ele-W})
become mass terms for elementary fields in the dual $SU(2)\times
SU(2)$ theory. By integrating out these fields, we arrive at the
following theory.
\begin{table}[htbp]
  \begin{center}
    \leavevmode $
\begin{array}{|c|cc|ccccc|} \hline
  & SU(2)_1 & SU(2)_2 & SU(4) & SU(2N+1) & U(1)_1 & U(1)_2 & U(1)_R \\
  \hline
  q & \f & 1 & \fb & 1 & -\frac{N}{2} & 0 & 0 \\
  x & \f & \f & 1 & 1 & N & 0 & 1 \\
  y & 1 & \f & 1 & \fb & 0 & 1 & 1-\frac{1}{N} \\
  l & 1 & \f & 1 & 1 & -2N & -2N-1 & -1+\frac{1}{N} \\
  M & 1 & 1 & \f & \f & -\frac{N}{2} & -1 & \frac{1}{N} \\
  M' & 1 & 1 & 1 & \f & 0 & 2N & 0 \\
  P & 1 & 1 & 1 & \asym & 0 & -2 & \frac{2}{N} \\
  P' & 1 & 1 & \f & 1 & \frac{3}{2}N & 2N+1 & 2-\frac{1}{N} \\ \hline
\end{array} $
    \caption{The other dual of the electric theory.}
    \label{tab:dualofle}
  \end{center}
\end{table}
The superpotential is
\begin{eqnarray}
  W = q^4 x^2 +Mqxy +M'ylx^2 +Py^2 +P'qxl +P^NM'.
  \label{dual-W}
\end{eqnarray}
Under this dual transformation, the mapping of the gauge invariant
operators in the chiral ring is
\begin{eqnarray}
  \begin{array}{c}
    ~~V'\overline{Q} \leftrightarrow M, \qquad Q''\overline{Q}
    \leftrightarrow M', \qquad V\overline{Q}^2 \leftrightarrow P,
    \\[1mm]
    V^{N-1}V'Q'' \leftrightarrow P', \qquad V^N \leftrightarrow x^2,
    \qquad \overline{Q}^{2N} \leftrightarrow yl.
  \end{array}
\end{eqnarray}
Hereafter we omit various scale factors which appear in the
superpotential and duality mapping for simplicity. In the above
superpotential (\ref{dual-W}), we added the last term which is allowed
by all the flavor and gauge symmetries and holomorphy resulting in
a superpotential which is the most generic one. This type of terms is
important in considering the relevant description of the low-energy
physics. Generally, in confining theories, its existence may be
interpreted as the non-perturbative (instanton) effects in
the completely broken dual gauge group. It should be noted that even
when there are terms allowed by symmetries and holomorphy arguments in
the deconfined theories, they are not generated in the low-energy
theories unless they can be nonsingularly composed by the products of
composite operators.

Then, the $SU(2)_1$ theory has three flavor quarks and smoothly
confines. After the confinement of the $SU(2)_1$ theory, finally, the
field content of the resultant dual theory is given in Table
\ref{tab:dual}
\begin{table}[htbp]
  \begin{center}
    \leavevmode $
\begin{array}{|c|c|ccccc|} \hline
  & SU(2)_2 & SU(4) & SU(2N+1) & U(1)_1 & U(1)_2 & U(1)_R \\ \hline
  y & \f & 1 & \fb & 0 & 1 & 1-\frac{1}{N} \\
  l & \f & 1 & 1 & -2N & -2N-1 & -1+\frac{1}{N} \\
  M & 1 & \f & \f & -\frac{N}{2} & -1 & \frac{1}{N} \\
  M' & 1 & 1 & \f & 0 & 2N & 0 \\
  P & 1 & 1 & \asym & 0 & -2 & \frac{2}{N} \\
  P' & 1 & \f & 1 & \frac{3}{2}N & 2N+1 & 2-\frac{1}{N} \\
  T=[qq] & 1 & \asym & 1 & -N & 0 & 0 \\
  T'=[qx] & \f & \fb & 1 & \frac{N}{2} & 0 & 1 \\
  T''=[xx] & 1 & 1 & 1 & 2N & 0 & 2 \\ \hline
\end{array} $
    \caption{The low-energy description of a dual of the electric
      theory.}
    \label{tab:dual}
  \end{center}
\end{table}
and the superpotential is
\begin{eqnarray}
  W = T^2T'' +TT'^2 +MT'y +M'T''yl +Py^2 +P'T'l +P^NM'.
\end{eqnarray}

On the other hand, the $Sp(2N)$ magnetic theory also has a known dual
description.  In this dual picture, again, some of the terms in the
superpotential (\ref{exa-mag-W}) become mass terms and some fields are
decoupled from the low-energy dynamics. Integrating out these fields,
we have the matter content shown in Table \ref{tab:other-dual},
\begin{table}[htbp]
  \begin{center}
    \leavevmode $
\begin{array}{|c|c|ccccc|} \hline
  & Sp(2) & SU(4) & SU(2N+1) & U(1)_1 & U(1)_2 & U(1)_R \\ \hline
  m & \f & 1 & \fb & 0 & 1 & 1-\frac{1}{N} \\
  b' & \f & 1 & 1 & -2N & -2N-1 & -1+\frac{1}{N} \\
  q' & \f & \fb & 1 & \frac{N}{2} & 0 & 1 \\
  M' & 1 & 1 & \f & 0 & 2N & 0 \\
  B & 1 & 1 & 1 & 2N & 0 & 2 \\
  Z & 1 & 1 & \asym & 0 & -2 & \frac{2}{N} \\
  Z' & 1 & \f & \f & -\frac{N}{2} & -1 & \frac{1}{N} \\
  Z'' & 1 & \f & 1 & \frac{3}{2}N & 2N+1 & 2-\frac{1}{N} \\
  Z''' & 1 & \asym & 1 & -N & 0 & 0 \\ \hline
\end{array} $
    \caption{The other dual of the magnetic theory.}
\label{tab:other-dual}
  \end{center}
\end{table}
and the superpotential is given by
\begin{eqnarray}
  W = Z^NM' +M'Bmb' +Zm^2 +Z'mq' +Z''q'b' +Z'''q'^2 +BZ'''^2.
\end{eqnarray}
We also added the last term which is allowed by all symmetries
and holomorphy, and which may concern the non-perturbative dynamics in
the broken dual gauge group. The above superpotential is the most
general one with respect to all the symmetries and holomorphy. The
operator mapping in the chiral ring between the magnetic and dual
theories is as follows:
\begin{eqnarray}
  M^2 \leftrightarrow Z, \quad MQ' \leftrightarrow Z', \quad B'Q'
  \leftrightarrow Z'', \quad Q'^2 \leftrightarrow Z'''.
\end{eqnarray}
The gauge singlet meson $[MB']$ is decoupled by the mass term with
$\overline{B}$\@. Remarkably enough, this theory is the same as the
above $SU(2)_2$ theory which is the dual of the electric $SU(2N)$
theory. The one-to-one correspondence of the elementary fields are
easily seen from the tables:
\begin{eqnarray}
  \begin{array}{ccc}
    y \leftrightarrow m,\quad & l \leftrightarrow b',\quad & M
    \leftrightarrow Z', \\[1mm]
    M' \leftrightarrow M',\quad & P \leftrightarrow Z,\quad & P'
    \leftrightarrow Z'', \\[1mm]
    T \leftrightarrow Z''',\quad & T' \leftrightarrow q',\quad & T''
    \leftrightarrow B.
  \end{array}
\end{eqnarray}
{}From this, we can explicitly confirm that the electric and magnetic
theories are certainly dual.

It should be noted that the above dual pair gives a new example of
chiral to non-chiral dualities. So far, this type of duality has
been obtained only in the restricted type of theories; that is, the
dualities between $SO$ gauge theory with spinors in the real
representation \cite{pouliot2,pouliot3,cho} and $SU$ gauge theory with
a second rank symmetric tensor, which is a chiral theory. In addition,
those chiral to non-chiral dualities are highly nontrivial and less
easily found. However, it is interesting that with our method, we can
easily discover many new dualities of this type by only setting
product gauge groups as one wants. More examples of new duals of this
type are presented below.

We show the next example of dualities constructed by using our
method. As with the previous example, we can check the consistency
of the duality in a nontrivial manners which incorporate the
non-perturbative dynamics. However, we here concentrate on describing
new dual pairs. Instead of the $SU(2N)$ gauge group, we can consider
$SU(2N-1)$ theory which may be interesting from phenomenological
points of view. The matter content of the deconfined theory which we
now consider is shown in Table \ref{tab:odd-exa}, and the tree-level
superpotential $W=0$\@.
\begin{table}[htbp]
  \begin{center}
    \leavevmode $
\begin{array}{|c|cc|cccc|} \hline
     & SU(2N-1) & Sp(2N) & SU(5) & SU(2N) & U(1) & U(1)_R \\ \hline
  Q  &  \f  &  \f  &  1  &  1  & \frac{-5}{2N-1} & \frac{1}{N+2} \\
  Q' &  1  &  \f  &  \f   &  1  &  1  & \frac{1}{N+2} \\
  \overline{Q} & \fb &  1  &  1  &  \f  & \frac{5}{2N-1} &
  \frac{2}{N(N+2)} \\ \hline
\end{array} $
    \caption{The deconfined theory.}
    \label{tab:odd-exa}
  \end{center}
\end{table}

We can obtain a dual pair by confining each gauge group in this
example as well as in the previous one. After the $Sp(2N)$ group
confines, we have the electric $SU(2N-1)$ theory and the matter
content is given in Table \ref{tab:odd-ele}.
\begin{table}[htbp]
  \def\arraystretch{1.1}
  \begin{center}
    \leavevmode $
\begin{array}{|c|c|cccc|} \hline
     & SU(2N-1) & SU(5) & SU(2N) & U(1) & U(1)_R \\ \hline
  V=[QQ] &  \asym  &  1  &  1  & \frac{-10}{2N-1} & \frac{2}{N+2} \\
  V'=[QQ'] &  \f  &  \f  &  1  & \frac{2N-6}{2N-1} & \frac{2}{N+2} \\
  V''=[Q'Q'] &  1  &  \asym  &  1  &  2  & \frac{2}{N+2} \\
  \overline{Q} & \fb &  1  &  \f  & \frac{5}{2N-1} & \frac{2}{N(N+2)}
  \\ \hline
\end{array} $
    \caption{The electric theory.}
  \label{tab:odd-ele}
  \end{center}
\end{table}
This theory has a tree-level (dynamical generated) superpotential
\begin{eqnarray}
  W = V^{N-1}V'V''^2 +V^{N-2}V'^3V'' +V^{N-3}V'^5.
\end{eqnarray}
On the other hand, by confining the $SU(2N-1)$ theory at first, we
have the magnetic theory as shown in Table \ref{tab:odd-mag}
\begin{table}[htbp]
  \def\arraystretch{1.1}
  \begin{center}
    \leavevmode $
\begin{array}{|c|c|cccc|} \hline
     & Sp(2N) & SU(5) & SU(2N) & U(1) & U(1)_R \\ \hline
  M=[Q\overline{Q}] &  \f  &  1  &  \f  &  0  & \frac{1}{N} \\
  B=[Q^{2N-1}] &  \f  &  1  &  1  &  -5  & \frac{2N-1}{N+2} \\
  \overline{B}=[\overline{Q}^{2N-1}] &  1  &  1  &  \fb  &  5  &
  \frac{4N-2}{N(N+2)} \\
  Q' &  \f  &  \f   &  1  &  1  & \frac{1}{N+2} \\ \hline
\end{array} $
    \caption{The magnetic theory.}
    \label{tab:odd-mag}
  \end{center}
\end{table}
with the superpotential
\begin{eqnarray}
  W = M^{2N} +BM\overline{B}.
\end{eqnarray}
This is also an example of chiral to non-chiral duality based on
simple gauge groups.

Another interesting feature of this method for finding dualities is
that the mapping of gauge invariant operators is constructed very
easily. This can be done by considering gauge invariant operators
in the deconfined theory. By decomposing gauge invariant operators in
both electric and magnetic theories, we can see the operator mapping
between the chiral rings of those theories without comparing the
transformation properties under flavor symmetries.

In addition, these dual pairs are also certainly stable with
deformations of two theories as well as the known dual pairs. By some
deformations, it may occur that one theory flows out of the
non-abelian Coulomb phase and at the same time the dual gauge group
does not completely broken. But in this situation, we can also expect
that two theories describe the same infrared physics; confinement,
chiral symmetry breaking, dynamical supersymmetry breaking, etc. We
will actually see such deformations in the next section.

We can find many other dual pairs based on other gauge groups by
altering the way to gauge flavor symmetries in various s-confining
theories. For instance, we can straightforwardly construct dual
pairs of theories based on the gauge groups $SU(2N)$, $SU(2N-1)$
versus $Sp(2N-2)$, $Sp(2N-4)$ as in the previous examples. Especially,
the magnetic $Sp(2N-4)$ theory (versus $SU(2N)$ or $SU(2N-1)$) can be
found from the known duality which is obtained by the deconfinement
method \cite{berkooz,pouliot}. We can obtain these dualities by
setting the number of flavors in the electric theory to the critical
one such that one of the dual gauge groups, which is of product form,
becomes
trivial and the remaining dual gauge group is simple. In this way,
some of the dual pairs obtained by our method can be interpreted as
the critical situations of the known dualities obtained by the
deconfinement method. However, among theories which have more flavors
than s-confining theories, there are only a few theories whose magnetic
descriptions have been known. Therefore, our method may be a more
powerful tool than the usual deconfinement method in order to find new
dual pairs of theories. Conversely, we may construct new dual theories
based on the ``product'' gauge group as the duals of theories which
contain more fields than the electric theories obtained by our method.

%%%%%%%%%%%%%%%%%%%%%%%%%%%%%%%%%%%%%%%%%%%%%%%%%%%%%%%
%%%%%%%%%%%%%%%%%%%%%%%%%%%%%%%%%%%%%%%%%%%%%%%%%%%%%%%
%%%%%---------------y  section  z---------------%%%%%
\section{Model and application}
\setcounter{equation}{0}
\setcounter{table}{0}

\subsection{Dynamical supersymmetry breaking}

Dynamical supersymmetry breaking is a very interesting phenomenon for
theoretical and phenomenological meanings and has been intensively
investigated for a long time \cite{DSB}. In chiral gauge theories,
many models of dynamical supersymmetry breaking have been
constructed. In these models, there is no classical flat direction and
some of the global symmetries are broken by strong coupling
dynamics. That
is, by the quantum effects, supersymmetry is broken dynamically. Among
these theories, one famous model is based on $SU(5)$ gauge theory with
an antisymmetric tensor and an antifundamental field \cite{su5}. This
model has no gauge invariant operator, so the classical moduli space
is only at the origin of the field space. Since this model is
asymptotically free, at the origin it is very strongly coupled
in the infrared region and it is plausible that global ($U(1)_R$)
symmetry is broken by non-perturbative dynamics. Therefore,
supersymmetry is considered to be broken according to the argument of
the Nambu-Goldstone theorem or the Konishi anomaly \cite{konishi}. So
far, this probable supersymmetry breaking has been confirmed by
considering the flavor decoupling from calculable models
\cite{murayama} or the weakly coupled magnetic descriptions of models
which contain more fundamental flavors than this $SU(5)$ model
\cite{pouliot}. But in this dual description, the magnetic gauge group
is completely broken through the Higgs mechanism
or confined and then becomes trivial due to the
effects corresponding to the mass decoupling effects. So, this
argument is only equivalent to considering the flavor decoupling from
the s-confining theory in the electric side, which is only a
Wess-Zumino model for the composite gauge singlets. In this
subsection, we present another check of this supersymmetry breaking
scenario by directly analyzing the non-abelian dual description of
this $SU(5)$ model. A similar analysis has been done in the different
context \cite{luty2}. As is seen below, though this dual theory has
non-chiral field content, the dynamical supersymmetry breaking indeed
occurs when taking into account the non-perturbative effects
properly.

Let us consider the following deconfined theory with zero tree-level
superpotential (Table \ref{tab:dsb-deconf}).
\begin{table}[htbp]
  \def\arraystretch{1.1}
  \begin{center}
    \leavevmode $
\begin{array}{|c|cc|ccccc|} \hline
    & SU(5) & SU(4) & SU(5) & SU(5) & U(1)_1 & U(1)_2 & U(1)_R \\ \hline
  A & \asym & 1 & 1 & 1 & 1 & 4 & -6/5 \\
  Q & \f & \f & 1 & 1 & 0 & -3 & -3/5 \\
  \overline{Q} & \fb & 1 & \f & 1 & -3/5 & 0 & 8/5 \\
  Q' & 1 & \fb & 1 & \fb & 0 & 3 & 1 \\ \hline
\end{array} $
    \caption{The deconfined theory.}
    \label{tab:dsb-deconf}
  \end{center}
\end{table}
Both gauge groups of $SU(5)$ and $SU(4)$ confine at $\Lambda_1$ and
$\Lambda_2$, respectively. After confinement of each gauge group at
each limit, $\Lambda_1 \ll \Lambda_2$ or $\Lambda_1 \gg \Lambda_2$, we
have the following dual pair of theories. The field content of the
electric theory is shown in Table \ref{tab:dsb-ele} and a
dynamically generated superpotential is given by
\begin{table}[htbp]
  \begin{center}
    \leavevmode $
\begin{array}{|c|c|ccccc|} \hline
       & SU(5) & SU(5) & SU(5) & U(1)_1 & U(1)_2 & U(1)_R \\ \hline
  M=[QQ'] & \f & 1 & \fb & 0 & 0 & 2/5 \\
  B=[Q^4] & \fb & 1 & 1 & 0 & -12 & -12/5 \\
  \overline{B}=[Q'^4] & 1 & 1 & \f & 0 & 12 & 4 \\
  A & \asym & 1 & 1 & 1 & 4 & -6/5 \\
  \overline{Q} & \fb & \f & 1 & -3/5 & 0 & 8/5 \\ \hline
\end{array} $
    \caption{The electric theory.}
    \label{tab:dsb-ele}
  \end{center}
\end{table}
\begin{eqnarray}
  W = M^5 +BM\overline{B}.
\end{eqnarray}
The magnetic theory becomes as shown in Table \ref{tab:dsb-mag} and has
a superpotential
\begin{table}[htbp]
  \def\arraystretch{1.1}
  \begin{center}
    \leavevmode $
\begin{array}{|c|c|ccccc|} \hline
    & SU(4) & SU(5) & SU(5) & U(1)_1 & U(1)_2 & U(1)_R \\ \hline
  V=[Q\overline{Q}] & \f & \f & 1 & -3/5 & -3 & 1 \\
  P=[A\overline{Q}^2] & 1 & \asym & 1 & -1/5 & 4 & 2 \\
  R=[A^2Q] & \f & 1 & 1 & 2 & 5 & -3 \\
  T=[AQ^3] & \fb & 1 & 1 & 1 & -5 & -3 \\
  U=[\overline{Q}^5] & 1 & 1 & 1 & -3 & 0 & 8 \\
  Q' & \fb & 1 & \fb & 0 & 3 & 1 \\ \hline
\end{array} $
    \caption{The magnetic theory.}
    \label{tab:dsb-mag}
  \end{center}
\end{table}
\begin{eqnarray}
  W = V^3PR + VP^2T + RTU.
\end{eqnarray}
Here let us consider the deformation of this duality. This can be
done by adding a tree-level superpotential in the deconfined
theory. We add the following superpotential:
\begin{eqnarray}
  W = \sum_{a,i=1}^{5} h^{ai}Q\overline{Q}_aQ'_i.
  \label{treeW}
\end{eqnarray}
With this term, the flavor symmetries $SU(5)\times SU(5)\times
U(1)_1$ are broken to $SU(5)_{\rm diag}$\@. This term becomes the mass
term
of the elementary fields in both theories after confinements. In the
electric $SU(5)$ theory, integrating out these massive modes, we
arrive at the $SU(5)$ model with matter fields consisting one
antisymmetric tensor and one antifundamental tensor. Since this
theory has
no gauge invariant operator, we also have no superpotential. This is
just the above-mentioned theory (except for a singlet field
$\overline{B}$ which has no interaction and is physically irrelevant
in this case)\@. In this theory supersymmetric vacuum is probably
lifted by the $SU(5)$ strong dynamics.

On the other hand, after we integrate out the massive fields, the
low-energy description of the magnetic theory is an $SU(4)$ gauge
theory with one flavor quarks and singlet fields with the following
superpotential;
\begin{table}[htbp]
  \begin{center}
    \leavevmode $
\begin{array}{|c|c|cc|} \hline
    & SU(4) &SU(5)& U(1)_R \\ \hline
  P & 1 & \asym& 2 \\
  R & \f & 1& -3 \\
  T & \fb & 1& -3 \\
  U & 1 & 1& 8 \\ \hline
\end{array} $
    \caption{The deformed magnetic theory.}
    \label{tab:dsb-mag-def}
  \end{center}
\end{table}
\begin{eqnarray}
  W = RTU + 3\left(\frac{\widetilde{\Lambda}_2^{11}}{RT}\right)^{1/3},
\end{eqnarray}
where $\widetilde{\Lambda}_2$ is the dynamical scale of the above
$SU(4)$ theory (Table \ref{tab:dsb-mag-def})\@. The second term is
induced by the non-perturbative effects in the $SU(4)$ theory. Without
the second term this model has a classical flat direction associated
with the singlet field $U$\@. (The flat direction along $P$ is not
concerned with any dynamics, so we neglect this direction in the
following.) However, as seen in the following, this flat direction is
stabilized in the full quantum theory. For this purpose, it is
convenient to consider the theory with $U$ as a fixed parameter
\cite{shirman}. In this theory, clearly supersymmetry is unbroken and
the meson $m=RT$ has a vacuum expectation value  $m\sim U^{-3/4}$ in
the vacuum. Therefore
we find a runaway property of the full theory potential $V$\@:
\begin{eqnarray}
  V = \left|\frac{\partial W}{\partial U}\right|^2 \,\sim\,
  U^{-3/2}.
\end{eqnarray}
However, it is noted that this picture is valid only for $U \ll
\Lambda_1$\@, i.e., only when $U$ is sufficiently smaller than
the $SU(5)$ confinement scale. In the case of $U \gg \Lambda_1$, the
relevant description is the deconfined theory. In this picture, the
classical flat direction is parametrized by
$U=\ol{Q}^5=v^5$\@. Reconsidering the behavior of the potential, we
find
\begin{eqnarray}
  V = \left|\frac{\partial W}{\partial \ol{Q}}\right|^2 \,\sim\,
  U^{-3/2} v^{4\cdot 2} \,\sim\, v^{1/2}.
\end{eqnarray}
Thus the classical flat direction is stabilized by the quantum
effects. In the finite size region of the moduli space, supersymmetry
is surely broken by the contradictory F-flatness conditions of $U$ and
$m$\@.

In this way, we can directly confirm the probable dynamical
supersymmetry breaking scenario of the $SU(5)$ chiral model in terms
of its non-chiral dual description in a consistent manner by using the
non-perturbative dynamics. Similar analyses can be made for the
$SU(2N+1)$ theory with one antisymmetric tensor and $(2N-3)$
antifundamental fields. With the relevant tree-level terms to lift
the classical flat direction, it is known that this theory has a
stable supersymmetry breaking vacuum \cite{su-odd,pouliot}. We can
also check this supersymmetry breaking phenomenon by considering the
dual gauge dynamics obtained in the previous section which has a
non-abelian $Sp$ gauge group and non-chiral field content.

%%%%%%%%%%%%%%%%%%%%%%%%%%%%%%%%%%%%%%%%%%%%%%%%%%%%%%%
%%%%%%%%%%%%%%%%%%%%%%%%%%%%%%%%%%%%%%%%%%%%%%%%%%%%%%%
%%%%%---------------y  section  z---------------%%%%%
\subsection{New confining theories}

In this section we present an example of new confining theories
with a tree-level superpotential as an application of our method. For
supersymmetric gauge theories without tree-level superpotential, it is
known that we can classify the sort of theories according to the
Dynkin index arguments. Especially, any theories which reveal the
s-confinement behavior have matter contents satisfying the index
constraint $\sum \mu_i=\mu_G+2$ and have been already listed
completely \cite{csaki}. This index condition severely restricts the
varieties of possible s-confining theories, consequently the
possibilities of their applications for model building. However, we
do not know what kind of theory confines in the presence of a 
tree-level superpotential. In this section, we show that some
examples of this type of confining model can be obtained by adding the
relevant superpotential to the deconfined theories so that one of the 
electric and magnetic theories can flow to the known confining
theories when taking into account the effects of this
perturbation. A similar type of confining theories has recently been
obtained \cite{new-sconf} by considering the case when the dual gauge
group reduces to trivial. New confining theories obtained in this
section are interpreted as the same type of theory as theirs, since
our method can be regarded as considering the cases in which the dual
gauge group reduces to a trivial one by confinement dynamics and/or
the Higgs mechanism.

We proceed the analysis by taking the $SU(5)\times SU(4)$ model in the 
previous section as an example, for simplicity. We use the same
notation for the matter fields as the previous one. Let us consider
adding the following perturbative superpotential in the deconfined
theory:
\begin{eqnarray}
  W = A^2QQ'_5 + A\ol{Q}^2X,
  \label{deformation}
\end{eqnarray}
where $Q'_5$ is the fifth component of $Q'$ and we introduced a new
field $X$ which is a gauge singlet but has the indices of the flavor
symmetries. As one can see below, the field $X$ is needed to guarantee
that the low-energy magnetic theory has no tree-level superpotential
and surely confines. This type of singlet field is generally used in
the deconfinement method (see Section 2). By the first term one of the
flavor $SU(5)$ symmetry is broken to $SU(4)$ but other symmetries are
not broken (by changing the $U(1)$ quantum number assignment of the
field $Q'$).

First, we see the magnetic $SU(4)$ theory deformed by these
perturbations. In this theory, these two terms in Eq.\
(\ref{deformation}) correspond to the mass terms of $P,R,Q'_5$ and
$X$\@. By integrating out these fields we have the following matter
content as shown in Table \ref{tab:conf-mag} ($\widehat{Q}'$ denotes
the first four components of $Q'$) with the zero tree-level
superpotential.
\begin{table}[htbp]
  \def\arraystretch{1.1}
  \begin{center}
    \leavevmode $
\begin{array}{|c|c|ccccc|} \hl
    & SU(4) & SU(5) & SU(4) & U(1)_1 & U(1)_2 & U(1)_R \\ \hl
  V=[Q\ol{Q}] & \f & \f & 1 & -3/5 & -3 & 1 \\
  T=[AQ^3] & \fb & 1 & 1 & 1 & -5 & -3 \\ 
  U=[\ol{Q}^5] & 1 & 1 & 1 & -3 & 0 & 8 \\
  \widehat{Q}' & \fb & 1 & \fb & 1/2 & 5 & 0 \\ \hl
\end{array} $
    \caption{The deformed magnetic theory.}
    \label{tab:conf-mag}
  \end{center}
\end{table}
Exactly speaking, an additional $U(1)$ flavor symmetry is
enhanced. However, it rotates only $U$ field and therefore is
irrelevant to the following analysis. If one is anxious about this
fact, one can introduce one more singlet field into the deconfined
theory and can integrate out $U$ away from the low-energy physics. The
above resultant $SU(4)$ theory is a supersymmetric QCD with five
flavor quarks and then s-confines. We show the confining spectra in
Table \ref{tab:conf-low} and the dynamically generated superpotential
is given by
\begin{eqnarray}
  W = ZZ'^4 + FZ\ol{F} + FZ'\ol{F}'.
  \label{confW}
\end{eqnarray}
\begin{table}[htbp]
  \begin{center}
    \leavevmode $
\begin{array}{|c|ccccc|} \hl
  & SU(5) &  SU(4) & U(1)_1 & U(1)_2 & U(1)_R \\ \hl
  Z=[VT] & \f & 1 & 2/5 & -8 & -2 \\
  Z'=[V\widehat{Q}'] & \f & \fb & -1/10 & 2 & 1 \\
  F=[V^4] & \fb & 1 & -12/5 & -12 & 4 \\
  \ol{F}=[\widehat{Q}'^4] & 1 & 1 & 2 & 20 & 0 \\
  \ol{F}'=[\widehat{Q}'^3T] & 1 & \f & 5/2 & 10 & -3 \\
  U & 1 & 1 & -3 & 0 & 8 \\ \hl
\end{array} $
    \caption{The low-energy effective theory of the magnetic theory.}
    \label{tab:conf-low}
  \end{center}
\end{table}
This model certainly describes a low-energy limit of the deconfined
theory with the superpotential (\ref{deformation}). The 't Hooft
anomaly matching conditions and the equivalence of the quantum moduli
space are satisfied between two theories.

In this case, a new confining theory can be constructed by considering
the electric side. By taking into account the $SU(4)$ dynamics at
first in the deconfined theory, we have the following electric $SU(5)$
theory deformed by the perturbations terms (Table \ref{tab:conf-ele})

\begin{table}[htbp]
  \def\arraystretch{1.1}
  \begin{center}
    \leavevmode $
\begin{array}{|c|c|ccccc|} \hl 
       & SU(5) & SU(5) & SU(4) & U(1)_1 & U(1)_2 & U(1)_R \\ \hl
  A & \asym & 1 & 1 & 1 & 4 & -6/5 \\
  \ol{Q} & \fb & \f & 1 & -3/5 & 0 & 8/5 \\
  X & 1 & \asymb & 1 & 1/5 & -4 & 0 \\
  M=[Q\widehat{Q}'] & \f & 1 & \fb & 1/2 & 2 & -3/5 \\
  M'=[QQ'_5] & \f & 1 & 1 & -2 & -8 & 22/5 \\
  B=[Q^4] & \fb & 1 & 1 & 0 & -12 & -12/5 \\
  \ol{B}=[\widehat{Q}'^3Q'_5] & 1 & 1 & \f & -1/2 & 10 & 5 \\
  \ol{B}'=[\widehat{Q}'^4] & 1 & 1 & 1 & 2 & 20 & 0 \\ \hl
\end{array} $
    \caption{The deformed electric theory which is expected to
      confine.}
    \label{tab:conf-ele}
  \end{center}
\end{table}
\noindent with a tree-level superpotential;
\begin{eqnarray}
  W = M^4M' +BM\ol{B} +BM'\ol{B}' +A^2M' +A\ol{Q}^2X.
  \label{tree}
\end{eqnarray}

At a glance, this $SU(5)$ theory is complicated. However, we have
already known from the above argument that this theory must confine
since it can be found that $SU(5)$ gauge symmetry breaking does not
occur with the tree-level superpotential. One interesting feature of
this method is that the confining massless spectra which satisfy the
't Hooft anomaly matching conditions and their confining
superpotentials are already given through another different pass in
the above. That is, they can be fixed from the viewpoint of the
magnetic theory, and all we have to do is to translate them into the
electric theory by pursuing the operator mapping. This operator
mapping is also easily identified by decomposing the gauge invariant
operators into the preon fields (namely, into the deconfined
theory). The confining spectra can be written as follows in terms of
the $SU(5)$ theory;
\begin{eqnarray}
  \begin{array}{c}
    Z=[A\ol{Q}B], \qquad Z'=[\ol{Q}M], \qquad F=[\ol{Q}^4B], \\
    \ol{F}=\ol{B}', \qquad \ol{F}'=[AM^3], \qquad U=[\ol{Q}^5].
  \end{array}
  \label{composite}
\end{eqnarray}
Here we used the same notation as in Table \ref{tab:conf-low}. The
dynamically generated confining superpotential for these massless
composite states is also given by Eq.\ (\ref{confW}) as previously.

It should be noted that in general, the consistency against
deformations only says that quantum moduli spaces of two theories are
still equivalent under these deformations. With our method in which we
use smoothly confining theories without a superpotential, even the
classical moduli spaces of the electric and magnetic theories are 
identical to each other. However, when we add perturbation terms to
the superpotential, the equivalence of the classical moduli is no
longer guaranteed but of course the quantum moduli are still
equivalent. This situation certainly takes place in the above electric
$SU(5)$ theory between high- and low-energy descriptions. In the
$SU(5)$ theory, there exist some gauge invariant operators such as
$[A\ol{Q}^2]$, $[A^2M]$, $[A^2M']$, $[AM^2M']$, $[\ol{Q}M']$, $[MB]$,
$[M'B]$, $[M^4M']$, $\ol{B}$ and $X$ in addition to the above
composite states (\ref{composite})\@. Some of these operators do not
span the classical moduli space due to the presence of the tree-level
superpotential. However, there are surely remaining operators
describing the classical moduli space of vacua in addition to the 
moduli space spanned by the operators (\ref{composite}). That is, the
classical moduli space is not the same between the high-energy
electric theory (Table \ref{tab:conf-ele}) and the low-energy confined
theory (Table \ref{tab:conf-low}). Therefore this confinement dynamics
is not s-confinement to be exact. However, additional classical
flat directions are certainly stabilized quantum mechanically. Along
such directions, the tree-level superpotential may change the dynamics
of the electric theory into another phase. As a result of this
dynamics, these moduli are not massless degrees of freedom in the
low-energy physics as expected.

We can surely confirm these confining spectra and superpotentials from
another dual theory in a nontrivial way. This other dual theory has an 
$SU(2)\times SU(2)$ gauge group and is obtained by applying the 
deconfinement method \cite{terning} to the above electric theory and
by considering the tree-level superpotentials as perturbation
terms. The filed content of this dual theory is shown in Table
\ref{tab:conf-dual} and the superpotential is given by
\begin{table}[htbp]
  \begin{center}
    \leavevmode $
\begin{array}{|c|cc|ccccc|} \hl
  & SU(2)_1 & SU(2)_2 & SU(5) & SU(4) & U(1)_1 & U(1)_2 & U(1)_R
  \\ \hl
  q & \f & 1 & \fb & 1 & 1/10 & -2 & 0 \\
  q' & \f & 1 & 1 & 1 & -1/2 & 10 & 4 \\
  \tilde{q} & \f & 1 & 1 & \f & 0 & 0 & 1 \\
  \tilde{q}' & \f & 1 & 1 & 1 & 5/2 & 10 & -4 \\
  x & \f & \f & 1 & 1 & 0 & 0 & 1 \\
  p & \f & 1 & 1 & 1 & -5/2 & -10 & 4 \\
  l & 1 & \f & \f & 1 & -1/10 & 2 & 1 \\
  l' & 1 & \f & 1 & 1 & 1/2 & -10 & -3 \\
  M & 1 & 1 & \f & \fb & -1/10 & 2 & 1 \\
  M' & 1 & 1 & \f & 1 & -13/5 & -8 & 6 \\
  M'' & 1 & 1 & 1 & \fb & 1/2 & -10 & -3 \\
  M''' & 1 & 1 & 1 & 1 & -2 & -20 & 2 \\
  B_1 & 1 & 1 & 1 & \fb & 5/2 & 10 & -3 \\
  B_1' & 1 & 1 & 1 & 1 & 0 & 0 & 2 \\
  X & 1 & 1 & \asymb & 1 & 1/5 & -4 & 0 \\
  \ol{B} & 1 & 1 & 1 & \f & -1/2 & 10 & 5 \\
  \ol{B}' & 1 & 1 & 1 & 1 & 2 & 20 & 0 \\ \hl
\end{array} $
\caption{The another dual description of the electric $SU(5)$ theory.}
\label{tab:conf-dual} 
  \end{center}
\end{table}
\begin{eqnarray}
  W &=& Mq\tilde{q} + M'q\tilde{q}' +M''q'\tilde{q} +M'''q'\tilde{q}'
  +B_1p\tilde{q} +B_1'p\tilde{q}' \\[1mm] &&\qquad\qquad +xlq +xl'q'
  +Xl^2 +x^2 +M''\ol{B} +M'''\ol{B}' +B_1'. \nonumber
\end{eqnarray}
The tree-level superpotential (\ref{tree}) in the electric theory
induces mass terms for some fields in the dual theory which are the
classical moduli in the absence of this superpotential. After
integrating out these fields, the $SU(2)_2$ gauge theory has three
flavor quarks and then confines. As a result, we have an $SU(2)_1$
theory whose matter content is displayed in Table
\ref{tab:conf-dual2}.
\begin{table}[htbp]
  \def\arraystretch{1.1}
  \begin{center}
    \leavevmode $
\begin{array}{|c|c|ccccc|} \hl
  & SU(2)_1 & SU(5) & SU(4) & U(1)_1 & U(1)_2 & U(1)_R \\ \hl
  V=[ll] & 1 & \asym & 1 & -1/5 & 4 & 2 \\
  V'=[ll'] & 1 & \f & 1 & 2/5 & -8 & -2 \\
  q & \f & \fb & 1 & 1/10 & -2 & 0 \\
  q' & \f & 1 & 1 & -1/2 & 10 & 4 \\
  \tilde{q} & \f & 1 & \f & 0 & 0 & 1 \\
  \tilde{q}' & \f & 1 & 1 & 5/2 & 10 & -4 \\
  p & \f & 1 & 1 & -5/2 & -10 & 4 \\
  M & 1 & \f & \fb & -1/10 & 2 & 1 \\
  M' & 1 & \f & 1 & -13/5 & -8 & 6 \\
  B_1 & 1 & 1 & \fb & 5/2 & 10 & -3 \\
  B_1' & 1 & 1 & 1 & 0 & 0 & 2 \\
  X & 1 & \asymb & 1 & 1/5 & -4 & 0 \\ \hl
\end{array} $
\caption{The another dual description of the electric $SU(5)$ theory
  after taking into account the effects of the deformations.}
\label{tab:conf-dual2}
  \end{center}
\end{table}
The superpotential is
\begin{eqnarray}
  W = Mq\tilde{q} + M'q\tilde{q}' +B_1p\tilde{q} +B_1'p\tilde{q}'
  +Vq^2 +V'qq' +B_1' +VX.
\end{eqnarray}
The $SU(2)_2$ confining dynamics also generates mass terms for some
fields. Moreover, by the presence of the linear term $B_1'$, the
$SU(2)_1$ gauge symmetry is broken and then some fields become
massive. These massive modes just correspond to the classical moduli
in the electric $SU(5)$ theory which are absent in the low-energy
description (Table \ref{tab:conf-low}). Thus we finally arrive at
just the same theory as that in Table \ref{tab:conf-low} when we
include a superpotential term induced by the one-instanton effect in
the broken $SU(2)_1$ gauge group. In this way, in the product dual
gauge theory the effects of the tree-level perturbation terms ensure
that the extra classical moduli are surely decoupled from low-energy
physics and the low-energy theory is described by a Wess-Zumino
model. In other words, in the electric side, the finally obtained 
confining superpotential (\ref{confW}) produces the quantum
constraints satisfied by the massless composite states which arise
from taking into account the effects resulting from the $F$-flatness
conditions of the tree-level superpotential and other quantum
mechanics. We do not know the actual origin of this mechanics yet.

One more interesting feature is that the above confining $SU(5)$
theory is not restricted by the index constraint $\sum \mu_i=\mu_G+2$
(but $\sum \mu_i >\mu_G+2$) due to the presence of the tree-level
superpotential. This is a general feature of confining theories
constructed by this method as mentioned in Section 2. In this way, we
can construct this type of confining theory with various gauge
groups, matter contents, and their tree-level superpotential by
altering the deconfined theories (gauge and flavor symmetries) and
perturbation terms in various ways. The wide possibility of this kind
of confining theory may be interesting in considering
phenomenological models, such as grand unified models, composite
models of quarks and leptons, etc.

%%%%%%%%%%%%%%%%%%%%%%%%%%%%%%%%%%%%%%%%%%%%%%%%%%%%%%%
%%%%%%%%%%%%%%%%%%%%%%%%%%%%%%%%%%%%%%%%%%%%%%%%%%%%%%%
%%%%%---------------y  section  z---------------%%%%%
\subsection{More examples with various labels}

In this section we examine more examples of new duality in which
both theories have a simple gauge group by considering deconfined
theories with various types of labels. The low-energy effective
behavior of the product gauge theory are also discussed in 
Refs.\ \cite{product,intri}. According to Eq.\ (\ref{index}) of the
Dynkin index, we can discuss some feature of low-energy effective
theories with various labels.

%%%%%%%%%%%%%%%%%%%%%%%%%%%%%%%%%%%%%%%%%%%%%%%%%%%%%%%
\subsubsection{s-confinement $\times$ s-confinement}

One interesting application of our method is to investigate dualities
for exceptional gauge groups. It is known that the only exceptional
gauge group which can s-confine is $G_2$\@. By using this $G_2$ gauge
theory in the deconfined theory with the label ``s-confinement
$\times$ s-confinement'', we can find a dual description of an
exceptional $G_2$ gauge theory. The duality in the exceptional $G_2$
gauge theory with $N_f$ fundamental representation matter has been
discussed by Pouliot \cite{pouliot2} in terms of the $SU(N_f-3)$
chiral gauge theory. In this section we give a dual description
of the $G_2$ gauge theory with an adjoint representation field. The
matter content of the deconfined theory which we should now consider
is shown in Table \ref{tab:g2-deconf} and the tree-level
superpotential is zero.
\begin{table}[htbp]
  \begin{center}
    \leavevmode $
    \begin{array}{|c|cc|cc|} \hline
         & G_2 & Sp(4) & U(1) & U(1)_R \\ \hline
      Q  & \f & \f &  1 & 1/5 \\
      Q' & \f &  1 & -4 & 1/5 \\
      q  & 1  & \f & -7 & 3/5 \\ \hline
    \end{array} $
    \caption{The deconfined theory.}
    \label{tab:g2-deconf}
  \end{center}
\end{table}

We can easily obtain the electric and magnetic theories in the same
way as before. When we consider the $Sp(4)$ gauge dynamics first, we
obtain the electric $G_2$ gauge theory. The matter content of this
theory is given in Table \ref{tab:g2-ele}.
\begin{table}[htbp]
  \begin{center}
    \leavevmode $
    \begin{array}{|c|c|cc|c} \hline
           & G_2 & U(1) & U(1)_R \\ \hline
      \Phi=[QQ] & {\rm adj.} & 2 & 2/5 \\
      F=[QQ] & \f & 2 & 2/5 \\
      F'=[Qq] & \f & -6 & 4/5 \\
      Q' & \f & -4 & 1/5 \\ \hline
    \end{array} $
    \caption{The electric $G_2$ gauge theory.}
    \label{tab:g2-ele}
  \end{center}
\end{table}
The superpotential of the electric $G_2$ gauge theory is
\begin{eqnarray}
  W &=& F^3F' +F^2\Phi F' +F\Phi^2F' +\Phi^3F'.
\end{eqnarray}

On the other hand, when we consider the $G_2$ gauge dynamics before
$Sp(4)$ gauge dynamics should be considered, we obtain the magnetic
$Sp(4)$ theory. The matter content is shown in Table \ref{tab:g2-mag} 
\begin{table}[htbp]
  \begin{center}
    \leavevmode $
    \begin{array}{|c|c|cc|c} \hline
      & Sp(4) & U(1) & U(1)_R \\ \hline
      M =[Q^2] & {\rm adj.} & 2 & 2/5 \\
      M'=[QQ'] & \f & -3 & 2/5 \\
      M''=[Q'Q'] &1 & -8 & 2/5 \\
      A =[Q^3] & \f & 3 & 3/5 \\
      A'=[Q^2Q'] & \asym & -2 & 3/5 \\
      A''=[Q^2Q'] & 1 & -2 & 3/5 \\
      B =[Q^4] & 1 & 4 & 4/5 \\
      B'=[Q^3Q'] & \f & -1 & 4/5 \\
      q & \f & -7 & 3/5 \\ \hline
    \end{array} $
  \end{center}
  \caption{The magnetic $Sp(4)$ gauge theory.}
  \label{tab:g2-mag}
\end{table}
and the superpotential of the magnetic $Sp(4)$ gauge theory is given by
\begin{eqnarray}
  W &=& M^4M'' +M^3M'^2 +A'^2M^2 +A'A''M^2 +A''^2M^2 +AA'MM'+AA''MM'
  \nonumber \\
  && +A^2M'^2 +MB'^2 +M'BB' +M''BB +A'^2B +A''^2B
  +A'AB'+A''AB'. \nonumber \\
\end{eqnarray}

Let us consider the deformation of these theories. We add the
following superpotential in the deconfined theory:
\begin{eqnarray}
  W &=& QqQ'.
  \label{g2-tree}
\end{eqnarray}
This superpotential breaks $U(1)$ flavor symmetry and the remaining
flavor symmetry is R-symmetry. Then we obtain the superpotential of
the electric theory as follows:
\begin{eqnarray}
  W &=& F^3F' +F^2\Phi F' +F\Phi^2F' +\Phi^3F' +F'Q',
\end{eqnarray}
and this superpotential makes $F'$ and $Q'$ massive. After we
integrate out these massive modes using their equations of motion, we
have the following deformed electric theory which has a zero
tree-level superpotential.
\begin{table}[htbp]
  \begin{center}
    \leavevmode $
    \begin{array}{|c|c|c|} \hline
             & G_2 & U(1)_R \\ \hline
      \Phi=[QQ] & {\rm adj.} & 1/5 \\
      F=[QQ] & \f & 1/5 \\ \hline
    \end{array} $
    \caption{The deformed $G_2$ gauge theory.}
  \end{center}
\end{table}
This theory really has the flavor symmetries $U(1)\times
U(1)_R$. Therefore one $U(1)$ symmetry is missed. However, we expect
that the duality is preserved by this deformation.

On the other hand, we can calculate the deformed magnetic theory
including the superpotential (\ref{g2-tree}). In the end we obtain the 
deformed magnetic theory whose field content is shown in
Table \ref{tab:g2-mag-deform}
\begin{table}[htbp]
  \begin{center}
    \leavevmode $
    \begin{array}{|c|c|c|} \hline
        & Sp(4) & U(1)_R \\ \hline
      M & {\rm adj.} & 1/5 \\
      M'' & 1 & 6/5 \\
      A & \f & 3/10 \\
      A' & \asym & 4/5 \\
      A'' & 1 & 4/5 \\
      B & 1 & 2/5 \\
      B' & \f & 2/5 \\ \hline
    \end{array} $
    \caption{The deformed magnetic theory.}
    \label{tab:g2-mag-deform}
  \end{center}
\end{table}
and a superpotential is
\begin{eqnarray}
  W&=& M^4M'' +A'^2M^2 +A'A''M^2 +A''M^2 +MB'^2 +M''BB +A'^2B +A''^2B
  \nonumber \\ 
  && +A'AB' +A''AB'.
\end{eqnarray}
This theory is expected to be a dual description of the $G_2$ gauge
theory with one adjoint and one fundamental matter fields without
superpotential.

This duality can be interpreted as the critical situation of the 
duality obtained by the deconfinement method. In this case the $G_2$
gauge theory is realized in the moduli of the $Spin(7)$ gauge theory
with an adjoint and a spinor representation matters. If we give a VEV
to the spinor field, the $Spin(7)$ gauge group is broken to $G_2$ 
gauge group and the components of the spinor field except for the
moduli are absorbed by the Higgs mechanism. The adjoint representation
in $Spin(7)$ gauge group is decomposed into an adjoint and a
fundamental representation in $G_2$ gauge group. Therefore we obtain
the electric $G_2$ gauge theory with an adjoint and a fundamental
matters. The duality in this $Spin(7)$ gauge theory can be obtained by
the deconfinement method. When we consider the deconfined theory as
the adjoint matter field is replaced by a composite state, the
deconfined theory has $Spin(7)\times Sp(4)$ gauge group. The dual
description of this $Spin(7)$ gauge theory with a spinor and some
fundamental matters has been studied \cite{cho} and we obtain the
$SU(2)\times Sp(4)$ as a dual gauge group. In order to obtain a dual
description of the $G_2$ gauge theory, we need to deform the
$SU(2)\times Sp(4)$ dual theory by taking into account the effects
which correspond to giving a VEV to the spinor field in the electric
$Spin(7)$ gauge theory. In the course of this deformation, several
matter fields become massive and the $SU(2)$ gauge theory
confines. Finally, we can exactly obtain the deformed magnetic theory
in Table \ref{tab:g2-mag-deform}.

%%%%%%%%%%%%%%%%%%%%%%%%%%%%%%%%%%%%%%%%%%%%%%%%%%%%%%%
\subsubsection{s-confinement $\times$ runaway}

So far, we have discussed the deconfined theories with label
``s-confinement $\times$ s-confinement''. In this section we
consider the deconfined theories with other labels. Since the moduli
space is holomorphic under couplings and furthermore theories which
i-confine or have runaway behaviors can be obtained by mass
deformations of s-confining theories, we expect to have consistent
results in these cases when the relative ratio of two dynamical scales
is exchanged.

We first consider the low-energy effective description of the
deconfined theory with gauge group $G_1 \times G_2$ labeled by
``s-confinement $\times$ i-confinement''. When the $G_1$ gauge theory
confines (s-confinement) at first, the sum of indices of the resultant
$G_2$ gauge theory is more than $\mu_{G_2}+2$ from the inequality
(\ref{index}) and then the theory is expected to confine or have a
dual description which describes the same infrared behavior. On the
other hand, when first the $G_2$ gauge theory confines
(i-confinement), the sum of indices of the resultant theory is more
than $\mu_{G_1}+2$, and the resultant theory is similarly expected to
confine or have a dual description. Therefore we can expect that the
``s-confinement $\times$ i-confinement'' theory confines or has dual a
description which is more relevant to describe the low-energy
physics.

More nontrivial examples compared with the deconfined theories
labeled by ``s-confinement $\times$ i-confinement'' are theories
labeled by ``s-confinement $\times$ runaway'' without a tree-level
superpotential. When we first consider the non-perturbative dynamics
of the $G_2$ gauge group, the Affleck-Dine-Seiberg superpotential is
generated and a VEV of some field tends to go to infinity. Since
theories which have runaway behavior are obtained by mass deformations
of s-confining theories and we have explicitly seen the equivalence of
moduli space between the electric and magnetic theories obtained 
from the deconfined theory labeled by ``s-confinement $\times$
s-confinement'' by giving some examples, we expect that same
phenomenon is supposed to occur when we first consider the strong
dynamics of $G_1$ gauge group. When the $G_1$ gauge theory confines
(s-confinement), we obtain the magnetic theory in which the sum of
indices increased by more than 2. If we take the $G_2$ theory which
has runaway behavior with the sum of indices $\mu_{G_2}-2$ in the
deconfined theory (for example, the $SU(N)$ gauge theory with $(N-1)$
fundamental flavors), the $G_2$ gauge theory is expected to confine or
have a dual description. The above fact that a VEV of some field must
go to infinity in the electric $G_1$ theory tells that an expectation
value of some field in the magnetic side must go to infinity when we
solve the equations of motion at the low-energy limit though no
dynamics which induces the runaway behavior seems to occur. Let us
check this expectation by using the following example.

The field content of the deconfined theory is given in Table
\ref{tab:run-deconf} 
\begin{table}[htbp]
  \begin{center}
    \leavevmode $
    \begin{array}{|c|cc|ccc|} \hline
        & SU(4) & Sp(4) & SU(5) & U(1) & U(1)_R \\ \hline
      Q & \f & \f & 1 & 0 & -1/2 \\
      q & \f & 1 & 1 & -5 & 4 \\
      Q' & \fb & 1 & \f & 1 & 0 \\ \hline
    \end{array} $
    \caption{The deconfined theory.}
    \label{tab:run-deconf}
  \end{center}
\end{table}
and the tree-level superpotential is zero. The sum of indices of the
$SU(4)$ and the $Sp(4)$ gauge groups are $\mu_{SU(4)}+2$ and
$\mu_{Sp(4)}-2$, respectively.

Let us consider the $Sp(4)$ gauge dynamics first. In the electric
$SU(4)$ gauge theory, the Affleck-Dine-Seiberg superpotential is
generated and the VEV of $Q$ goes to infinity:
\begin{eqnarray}
  \VEV{Q}_a^{\alpha} &=& s\,\delta_a^{\alpha},\qquad s\rightarrow
  \infty, 
  \label{s}
\end{eqnarray}
where $\alpha$ and $a$ denote the index of the$SU(4)$ and $Sp(4)$
gauge group, respectively. Since we should consider the D-flatness
condition of $SU(4)$ gauge group, the form of $\VEV{Q}$ is
restricted to diagonal form. Along this direction, the product gauge
symmetry is broken to diagonal $Sp(4)$ gauge symmetry and the field
content of this electric theory is given in Table \ref{tab:run-ele}
where $U(1)'_R$ is the remaining R-symmetry which has no gauge anomaly.
\begin{table}[htbp]
  \begin{center}
    \leavevmode $
    \begin{array}{|c|c|ccc|} \hline
        & Sp(4) & SU(5) & U(1) & U(1)_R' \\ \hline
      q & \f & 1 & -5 & 0 \\
      Q' & \f & \f & 1 & 0\\ \hline
    \end{array} $
    \caption{The electric $Sp(4)$ theory.}
    \label{tab:run-ele}
  \end{center}
\end{table}

Since this $Sp(4)$ gauge theory has six fundamental fields, the theory
confines with chiral symmetry breaking (i-confinement) and becomes a
Wess-Zumino model. The matter content of the low-energy effective
theory is shown in Table \ref{tab:run-ele-low}
\begin{table}[htbp]
  \begin{center}
    \leavevmode $
    \begin{array}{|c|ccc|} \hline
          & SU(5) & U(1) & U(1)_R' \\ \hline
      F=[Q'q] & \f & -4 & 0 \\
      V=[Q'Q'] & \asym & 2 & 0 \\ \hline
    \end{array} $
    \caption{The low-energy description of the electric theory.}
    \label{tab:run-ele-low}
  \end{center}
\end{table}
and the superpotential is given by
\begin{eqnarray}
  W &=& X(FV^2-\Lambda_{D}^3),
\end{eqnarray}
where $X$ is an auxiliary field and $\Lambda_{D}$ is the dynamical
scale of the diagonal $Sp(4)$ gauge group. The constraint obtained
from the equation of the motion by $X$ breaks the global symmetry
$SU(5)$ to $Sp(4)$.

On the other hand, when we consider the $SU(4)$ gauge dynamics first
the $SU(4)$ gauge symmetry confines without chiral symmetry breaking
(s-confinement). The matter content of the magnetic $Sp(4)$ gauge
theory is given in Table \ref{tab:run-mag}
\begin{table}[htbp]
  \begin{center}
    \leavevmode $
    \begin{array}{|c|c|ccc|} \hline
      &Sp(4)&SU(5)&U(1)&U(1)_R \\ \hline
      M=[QQ'] & \f & \f & 1 & -1/2 \\
      F=[qQ'] & 1 & \f & -4 & 4 \\
      S=[Q^4] & 1 & 1 & 0 & -2 \\
      N=[Q^3q] & \f & 1 & -5 & 5/2 \\
      \ol{F}=[Q'^4] & 1 & \fb & 4 & 0 \\ \hline
    \end{array} $
    \caption{The magnetic $Sp(4)$ theory.}
    \label{tab:run-mag}
  \end{center}
\end{table}
and the superpotential is
\begin{eqnarray}
  W &=& M^4F +SF\ol{F} +NM\ol{F}.
\end{eqnarray}

As discussed before, the sum of indices increases by just 2 in this
case and the sum of indices of $Sp(4)$ gauge group in the magnetic
theory is equal to $\mu_{Sp(4)}$. Since this $Sp(4)$ gauge theory has
six fundamental matters, this theory confines with chiral symmetry
breaking (i-confinement) and the field content in the low-energy
effective description of the magnetic theory is given in Table
\ref{tab:run-mag-low}.
\begin{table}[htbp]
  \def\arraystretch{1.1}
  \begin{center}
    \leavevmode $
    \begin{array}{|c|ccc|} \hline
             & SU(5) & U(1) & U(1)_R \\ \hline
      A=[MM] & \asym & 2 & -1 \\
      P=[MN] & \f & -4 & 2 \\
      F & \f & -4 & 4 \\
      S & 1 & 0 & -2 \\
      \ol{F} & \fb & 4 & 0 \\ \hline
    \end{array} $
  \end{center}
  \caption{The low-energy description of the magnetic theory.}
  \label{tab:run-mag-low}
\end{table}
The superpotential is
\begin{eqnarray}
  W &=& A^2F +SF\ol{F} +P\ol{F} +X(A^2P-\Lambda_M^3),
\end{eqnarray}
where $X$ is an auxiliary field and $\Lambda_M$ is the dynamical scale
of the magnetic $Sp(4)$ gauge group. From the equations of motion we
find the runaway behavior
\begin{eqnarray}
  \VEV{S}\,\rightarrow\,\infty.
\end{eqnarray}
The superpotential makes a pair of $SU(5)$ fundamental and
antifundamental representation matter fields massive. In addition,
the constraint derived from the equation of motion by $X$ says that
the global symmetry $SU(5)$ is broken to $Sp(4)$.

In this way, we finally obtain the consistent results between the
electric and magnetic theories because the VEV $s$ in Eq.\ (\ref{s})
surely corresponds to $\VEV{S}$ in the magnetic theory, and the flavor
symmetries and the massless spectra are same between two
theories. Therefore, the expected runaway behavior is certainly
observed in the magnetic side as $\VEV{S}$ goes to infinity.

%%%%%%%%%%%%%%%%%%%%%%%%%%%%%%%%%%%%%%%%%%%%%%%%%%%%%%%
\subsubsection{affine-confinement $\times$ s-confinement}

As another type of confinement dynamics, it is known that there is the
affine confinement \cite{dotti1,dotti2} in which there is no
constraint between gauge invariant operators and therefore the
confining superpotential is not generated at the low-energy
region. Among the known affine-confining theories, the only candidate
having non-abelian flavor symmetries is the $SO(N)$ gauge theory with
$N-4$ vector representation matters. We try to consider the product
gauge theory using this $SO$ affine-confining theory.

The field content of the deconfined theory is displayed in Table
\ref{tab:aff-deconf}
\begin{table}[htbp]
  \begin{center}
    \leavevmode $
    \begin{array}{|c|cc|} \hline
        & SO(2N+4) & Sp(2N) \\ \hline 
      Q & \f & \f \\ \hline 
    \end{array} $
    \caption{The deconfined theory with the label ``affine-confinement
      $\times$ s-confinement''.}
    \label{tab:aff-deconf}
  \end{center}
\end{table}
and the tree-level superpotential is zero. This theory has no global
symmetry. After the $Sp(2N)$ confinement, the matter content of the
electric theory is given in Table \ref{tab:aff-ele}.
\begin{table}[htbp]
  \begin{center}
    \leavevmode $
    \begin{array}{|c|c|} \hline
      & SO(2N+4) \\ \hline
      \varphi=[QQ] & {\rm adj.} \\ \hline
    \end{array} $
    \caption{The electric theory.}
    \label{tab:aff-ele}
  \end{center}
\end{table}

\noindent This theory has a superpotential
\begin{eqnarray}
  W &=& \displaystyle{\frac{1}{\Lambda_{\rm sp}^{N-1}}}{\rm Pf}\,
  \varphi,
  \label{aff-W}
\end{eqnarray}
where $\Lambda_{\rm sp}$ is the dynamical scale of $Sp$ gauge group.

On the other hand, we first consider the $SO$ gauge dynamics. In this
case, there is a branch where the $SO(2N+4)$ gauge theory with $2N$
flavors confines and no dynamical superpotential is generated
(affine-confinement). Then the matter content of the magnetic theory
is given in Table \ref{tab:aff-mag}
\begin{table}[htbp]
  \begin{center}
    \leavevmode $
    \begin{array}{|c|c|} \hline
      & Sp(2N) \\ \hline
      \Phi=[QQ] & {\rm adj.} \\ \hline
    \end{array} $
    \caption{The magnetic theory}
    \label{tab:aff-mag}
  \end{center}
\end{table}
and the superpotential is zero. 

It is found that the degrees of freedom of the moduli space correspond
to each other. The gauge invariant operators of each theory are
written as follows:
\begin{eqnarray}
  \begin{array}{ccc}
    \mbox{electric} &~& \mbox{magnetic} \\[1mm]
    {\rm tr}\, \varphi^2, {\rm tr}\, \varphi^4, \cdots, {\rm tr}\,
    \varphi^{2N+2}, {\rm Pf}\, \varphi, &~& {\rm tr}\, \Phi^2, 
    {\rm tr}\, \Phi^4,\cdots, {\rm tr}\, \Phi^{2N}.
  \end{array}
\end{eqnarray}
The gauge invariant operators ${\rm tr}\, \varphi^{2N+2}$ and 
${\rm Pf}\, \varphi$ in the electric theory are set to be zero by the
equation of motion obtained from the dynamically generated
superpotential (\ref{aff-W}).

At a glance this magnetic $Sp(2N)$ gauge theory seems to be an 
${\cal N}=2$ supersymmetric theory. Of course, since we have little
information on the K\"ahlar potential, this theory may have only
${\cal N}=1$ supersymmetry. However, there may exist somewhere in the
moduli space where the low-energy effective theory becomes an 
${\cal N}=2$ supersymmetric theory. Similarly, the R-symmetry also
seems to be enhanced. We may understand this fact in the same way as
we understand the enhancement of supersymmetry.

It is interesting that if the magnetic theory has ${\cal N}=2$
supersymmetry in the infrared region, the above obtained duality
between the $SO$ and $Sp$ gauge theories is a duality between 
${\cal N}=1$ and ${\cal N}=2$ theories. In other words, the electric
$SO$ gauge theory with the superpotential (\ref{aff-W}) has 
${\cal N}=2$ supersymmetry in the infrared limit and the quantum
moduli space is the same as that in the $Sp$ magnetic theory.

%%%%%%%%%%%%%%%%%%%%%%%%%%%%%%%%%%%%%%%%%%%%%%%%%%%%%%%
%%%%%%%%%%%%%%%%%%%%%%%%%%%%%%%%%%%%%%%%%%%%%%%%%%%%%%%
%%%%%---------------y  section  z---------------%%%%%
\section{Summary and discussion}

In this paper, we have studied the ${\cal N}=1$ supersymmetric gauge
theories with product gauge group with various labels, especially
``s-confinement $\times$ s-confinement''. We can investigate the
behaviors of low-energy effective theories by using the facts that
the equivalence of moduli is valid for all value of two dynamical
scales and the inequality for the Dynkin index between the
elementary and the infrared matter fields. As a consequence, we can
find many dualities in theories which include the higher-rank tensor
representation fields. In this case, we generally obtain the dual
theories based on a simple gauge group, not a product one, and both
electric and magnetic theories have a tree-level
superpotential. Furthermore, we find that some of them are new
dualities between the chiral and non-chiral gauge theories. By giving
a few examples, we have justified this duality assumption in a
nontrivial way by using the known dualities and other strong gauge
dynamics. In some cases, the dual pairs obtained by our method can be
interpreted as the critical situations of the known dualities obtained
by the deconfinement method. However, among theories which have more
flavors than s-confining theories, there are only a few theories whose
magnetic descriptions have been known. Therefore, our method can be a
more powerful tool than the usual deconfinement method in order to
find new dual pairs. Conversely, we may use this duality assumption as
a hint to find dual theories with a product gauge group which contains
matter in the higher-rank tensor representations.

It is noted that the obtained duality is not the so-called strong-weak
duality. We expect that there may exist some other dual theory with
weak gauge coupling which is relevant to describe the low-energy
physics, and we have presented some examples.

It is interesting to consider some applications of new dualities 
by deforming these theories. With these deformations, in this paper we
have discussed the dynamical supersymmetry breaking phenomena, new
confining theories, and other interesting theoretical models. In every
case, we can see that the consistent behaviors between the two
theories are observed. New confining theories obtained in this paper
have several characteristic features. These theories are not
restricted by the constraint $\sum \mu_i =\mu_G+2$ because of the
existence of the tree-level superpotential. Moreover, some moduli in
the classical theories disappear in the infrared region. However, some
unknown physics may be needed to have a proper understanding of the 
consistency between two theories in studying new confining theories
and the affine-confinement dynamics. We leave these matters to future
investigations.

Moreover, the applications to phenomenological issues may also be
attractive. In this manner, we can construct various types of
dualities and confining theories. They generally contain higher-rank
tensor fields and have a nonzero superpotential. In addition, the
obtained new confining theories are not restricted by the index
argument and have wide possibilities for applying to model
building. It is interesting that one may find dual or more microscopic
descriptions of realistic models of our world.

\subsection*{Acknowledgements}

We would like to thank S.\ Sugimoto and M.\ Nitta for useful
discussions and comments. This work was supported in part by the
Grant-in-Aid for JSPS.

%%%%%%%%%%%%%%%%%%%%%%%%%%%%%%%%%%%%%%%%%%%%%%%%%%%%%%%
%%%%%%%%%%%%%%%%%%%%%%%%%%%%%%%%%%%%%%%%%%%%%%%%%%%%%%%
%%%%%--------------y reference z----------------%%%%%


\begin{thebibliography}{99}
%%%%% DSB
\bibitem{DSB} For a recent review, see \ E.\ Poppitz and S.P.\ Trivedi,
  hep-th/9803107.
%%%%% composite
\bibitem{composite} A.\ Nelson and M.J.\ Strassler, {\sl Phys.\ Rev.}
  {\bf D56} (1997) 4226;  M.A.\ Luty and R.N.\ Mohapatra, 
  {\sl Phys.\ Lett.} {\bf 396B} (1997) 161;  D.B.\ Kaplan, 
  F.\ Lepeintre and M.\ Schmaltz, {\sl Phys.\ Rev.} {\bf D56} (1997)
  7193; N.\ Kitazawa, hep-ph/9806229.
%%%%% duality
\bibitem{seiberg} N.\ Seiberg, {\sl Nucl.\ Phys.} {\bf B435} (1995)
  129, \\
  {\it ``Electric-Magnetic Duality in Supersymmetric Non-Abelian Gauge
    Theories''}
%%%%% deconfiment
\bibitem{berkooz} M.\ Berkooz, {\sl Nucl.\ Phys.} {\bf B452} (1995)
  513, \\
  {\it ``The Dual of Supersymmetric SU(2k) with an Antisymmetric
    Tensor and Composite Dualities''}
\bibitem{pouliot} P.\ Pouliot, {\sl Phys.\ Lett.} {\bf 367B} (1996) 151
\bibitem{luty} M.\ Luty, M.\ Schmaltz and J.\ Terning, 
  {\sl Phys.\ Rev.} {\bf D54} (1996) 7815
\bibitem{sakai} T.\ Sakai, {\sl Mod.\ Phys.\ Lett.} {\bf A12} (1997)
  1025 
\bibitem{su} W.C.\ Su, hep-th/9707076
\bibitem{terning} J.\ Terning, {\sl Phys.\ Lett.} {\bf 422B} (1998) 149
%%%%% Kutasov-Schwimmer
\bibitem{kutasov} D.\ Kutasov, {\sl Phys.\ Lett.} {\bf 351B} (1995)
  230, \\
  {\it ``A Comment on Duality in N=1 Supersymmetric Non-Abelian
    Gauge Theories''}
\bibitem{kutasov2} D.\ Kutasov and A.\ Schwimmer, {\sl Phys.\ Lett.}
  {\bf 354B} (1995) 315; 
  R.G.\ Leigh and M.J.\ Strassler, {\it ibid.} {\bf 356B} (1995) 492
%%%%% chiral to non-chiral (SO with a spinor)
\bibitem{pouliot2} P.\ Pouliot, {\sl Phys.\ Lett.} {\bf 359B} (1995)
  108, \\
  {\it ``Chiral Duals of Non-Chiral SUSY Gauge Theories''}
\bibitem{pouliot3} P.\ Pouliot and M.J.\ Strassler, {\sl Phys.\ Lett.}
  {\bf 370B} (1996) 76
\bibitem{pouliot4} P.\ Pouliot and M.J.\ Strassler, {\sl Phys.\ Lett.}
  {\bf 375B} (1996) 175
\bibitem{kawano} T.\ Kawano, {\sl Prog.\ Theor.\ Phys.} {\bf 95}
  (1996) 963
\bibitem{cho} P.\ Cho, {\sl Phys.\ Rev.} {\bf D56} (1997) 5260
%%%%% s-confiment
\bibitem{csaki} C.\ Cs\'aki, M.\ Schmaltz and W.\ Skiba, 
  {\sl Phys.\ Rev.} {\bf D55} (1997) 7840, \\
  {\it ``Confinement in N=1 SUSY Gauge Theories and Model Building
    Tools''}
%%%%% i- and c-quantum moduli
\bibitem{grinstein} B.\ Grinstein and D.R.\ Nolte, {\sl Phys.\ Rev.}
  {\bf D57} (1998) 6471; {\it ibid.} {\bf D58} (1998) 045012
%%%%% affine quantum moduli
\bibitem{dotti1} G.\ Dotti and A.\ Manohar, {\sl Phys.\ Rev.\ Lett.}
  {\bf 80} (1998) 2758, \\
  {\it ``Supersymmetric Gauge Theories With an Affine Quantum Moduli
    Space''}
\bibitem{dotti2} G.\ Dotti, A.\ Manohar and W.\ Skiba, 
  {\sl Nucl.\ Phys.} {\bf B531} (1998) 507
%%%%% product
\bibitem{product} K.\ Intriligator, R.G.\ Leigh and N.\ Seiberg, 
  {\sl Phys.\ Rev.} {\bf D50} (1994) 1092,\\
  {\it ``Exact Superpotentials in Four Dimensions''}\\
  E.\ Poppitz, Y.\ Shadmi and S.P.\ Trivedi, 
  {\sl Nucl.\ Phys.} {\bf B480} (1996) 125, \\
  {\it ``Duality and Exact Results in Product Group Theories''}
\bibitem{intri} K.\ Intriligator and S.\ Thomas, {\sl Nucl.\ Phys.}
  {\bf B473} (1996) 121; hep-th/9608046.
%%%%% index
\bibitem{manohar} G.\ Dotti and A.V.\ Manohar, {\sl Phys.\ Lett.}
  {\bf 410B} (1997) 167; {\sl Nucl.\ Phys.} {\bf B518} (1998) 575
%%%%% s-confine
\bibitem{s-conf} N.\ Seiberg, {\sl Phys.\ Rev.} {\bf D49} (1994) 6857;
  K.\ Intriligator and P.\ Pouliot, {\sl Phys.\ Lett.} {\bf 353B}
  (1995) 471 
%%%%% DSB
\bibitem{su5} I.\ Affleck, M.\ Dine and N.\ Seiberg, 
  {\sl Phys.\ Lett.} {\bf 137B} (1984) 187
\bibitem{konishi} K.\ Konishi, {\sl Phys.\ Lett.} {\bf 135B} (1984) 439
\bibitem{murayama} H.\ Murayama, {\sl Phys.\ Lett.} {\bf 355B} (1995)
  187;
  E.\ Poppitz and S.P.\ Trivedi, {\it ibid.} {\bf 365B} (1996) 125
\bibitem{luty2} M.A.\ Luty and J.\ Terning, {\sl Phys.\ Rev.} {\bf
  D57} (1998) 6799
\bibitem{shirman} Y.\ Shirman, {\sl Phys.\ Lett.} {\bf 389B} (1996)
  287
\bibitem{su-odd} I.\ Affleck, M.\ Dine and N.\ Seiberg, 
  {\sl Nucl.\ Phys.} {\bf B256} (1985) 557
%%%%% new-confinement
\bibitem{new-sconf} C.\ Cs\'aki and H.\ Murayama, {\sl Phys.\ Rev.}
  {\bf D59} (1999) 065001
\end{thebibliography}
\end{document}